\documentclass[conference]{IEEEtran}
\IEEEoverridecommandlockouts

\usepackage{amsmath,amssymb,amsfonts}
\usepackage[table]{xcolor}
\usepackage{siunitx}
\usepackage{comment}
\usepackage{adjustbox}
\usepackage{hyperref}
\usepackage{subcaption,graphicx}
\usepackage{soul}

\usepackage{tabularx}
\usepackage{multirow}
\usepackage{makecell}

\usepackage{pifont}
\newcommand{\cmark}{\ding{51}}
\newcommand{\xmark}{\ding{55}}

\usepackage[numbers]{natbib}

\graphicspath{{figures/}}

\usepackage{listings}
\definecolor{codegreen}{rgb}{0,0.6,0}
\definecolor{codegray}{rgb}{0.5,0.5,0.5}
\definecolor{codepurple}{rgb}{0.58,0,0.82}
\definecolor{backcolour}{rgb}{0.95,0.95,0.92}
\lstdefinestyle{mystyle}{
    backgroundcolor=\color{backcolour},   
    commentstyle=\color{codegreen},
    keywordstyle=\color{magenta},
    numberstyle=\tiny\color{codegray},
    stringstyle=\color{codepurple},
    basicstyle=\ttfamily\footnotesize,
    breakatwhitespace=false,                 
    captionpos=b,                    
    columns=fullflexible,
    keepspaces=true,
    showspaces=false,                
    showstringspaces=false,
    showtabs=false,                  
    tabsize=1
}
\title{Harnessing CUDA-Q's MPS for Tensor Network Simulations of Large-Scale Quantum Circuits\thanks{pre-print submitted for publication}}

\author{
\IEEEauthorblockN{Gabin Schieffer}
\IEEEauthorblockA{\textit{KTH Royal Institute of Technology} \\
Sweden \\
gabins@kth.se}
\and
\IEEEauthorblockN{Stefano Markidis}
\IEEEauthorblockA{\textit{KTH Royal Institute of Technology} \\
Sweden \\
markidis@kth.se}
\and
\IEEEauthorblockN{Ivy Peng}
\IEEEauthorblockA{\textit{KTH Royal Institute of Technology} \\
Sweden \\
ivybopeng@kth.se}}

\begin{document}

\maketitle

\begin{abstract}
Quantum computer simulators are an indispensable tool for prototyping quantum algorithms and verifying the functioning of existing quantum computer hardware. The current largest quantum computers feature more than one thousand qubits, challenging their classical simulators. State-vector quantum simulators are challenged by the exponential increase of representable quantum states with respect to the number of qubits, making more than fifty qubits practically unfeasible. A more appealing approach for simulating quantum computers is adopting the tensor network approach, whose memory requirements fundamentally depend on the level of entanglement in the quantum circuit, and allows simulating the current largest quantum computers. This work investigates and evaluates the CUDA-Q tensor network simulators on an Nvidia Grace Hopper system, particularly the Matrix Product State (MPS) formulation. We compare the performance of the CUDA-Q state vector implementation and validate the correctness of MPS simulations. Our results highlight that tensor network-based methods provide a significant opportunity to simulate large-qubit circuits, albeit approximately. We also show that current GPU-accelerated computation cannot fully utilize GPU efficiently in the case of MPS simulations.

\end{abstract}

\section{Introduction}
Quantum computing systems have been rapidly evolving in terms of capabilities, making the promise of fault-tolerant quantum computing closer day after day~\cite{pan_simulating_2021,tabuchi_mpiqulacs_2023,markidis2024quantum}. Today, two quantum computers, by IBM and Atom Computing, has broken the barrier of a thousand qubits. The rapid development of quantum computer hardware continuously simplifies scientists' access to actual quantum hardware, with an increased number of qubits, and improved error-correcting capabilities. However, while such access is technically possible, it still is limited by practical considerations. First, in the context of development and testing, the time and monetary cost needed to access quantum hardware can be a limiting factor, especially when compared to the high pace of algorithm developments. In addition, algorithms targeting future quantum computers, with a higher number of qubits, cannot be executed on real hardware. Therefore, scientists need to simulate the behavior of actual quantum hardware on classical computers to prototype, develop, and test their quantum algorithms.

Quantum simulation is a rich field of research, where several algorithms and optimizations have been proposed to simulate quantum hardware on classical computers. State vector simulation, where the full state vector is represented in memory, is ubiquitous and can simulate small-size quantum circuits, typically on the order of 30-40 qubits. However, this algorithm's memory footprint increases exponentially with the number of qubits, rendering it unpractical for high-qubits simulations, even with algorithmic optimizations. Works have been running optimized state vector simulations on large-scale compute clusters, for example, Fugaku supercomputer~\cite{tabuchi_mpiqulacs_2023}. However, the exponential memory footprint, which is inherent to the state vector method, creates a need for other simulation algorithms, when targeting the simulation of circuits with a higher number of qubits. 

Tensor network-based simulations have shown promising results in simulating high-qubit quantum circuits. Tensor network quantum circuit simulations are based on a compact and efficient quantum state representation in the form of factorized tensors, reconstructing the original quantum state vector. Tensor factorizations are not unique (therefore, different approaches exist) and can be exact or approximated (similar to \emph{lossy compression}). The tensor network approach allowed for quantum circuit simulation up to several hundreds of qubits~\cite{pan_simulating_2021,chen_cutn-qsvm_2024,nguyen_tensor_2022}. Furthermore, as tensor networks find applications in a wider field than quantum computing, this research area has gained significant traction in recent years. For instance, Nvidia leveraged the GPU-accelerated system of 896 GPUs to simulate the solution of the Max-Cut problem with 1,688 and 5,000 qubits in 2021. This shows that large-scale quantum computer simulations shy away from state vector and other more sophisticated techniques such as tensor networks need to be explored.%

We survey popular quantum frameworks for supporting tensor network simulations of quantum circuits. Quantum frameworks provide a set of tools, libraries, and algorithms to support domain scientists in developing quantum algorithms~\cite{qiskit_aer,cirq,faj2023quantum,azure_quantum}. These frameworks support seamless executions of quantum circuits, either on quantum hardware or on simulators running on classical computers. For instance, major quantum hardware vendors and actors provide quantum computer simulators that are executable on classical computing hardware, including simulators in IBM's Qiskit~\cite{qiskit_aer}, Google's Cirq~\cite{cirq} and qsim~\cite{qsim,markidis2023enabling}, Microsoft's Azure Quantum~\cite{azure_quantum}, or Nvidia's CUDA-Q~\cite{cudaq}. In this work, we focus on the CUDA-Q framework as it provides seamless integration with GPUs. 

We develop five quantum circuits representative of algorithms promising for solving complex real-world problems on quantum computers. We implement these circuits using the CUDA-Q framework and leverage the GPU-accelerated simulators to simulate the quantum circuits. In particular, we compare the tensor network-based simulation backend, including approximate matrix product state (MPS) simulation, with the state vector simulation backend for high counts of qubits to gain insights into the performance and scalability. Furthermore, we propose a method and an evaluation of the correctness of approximate tensor network-based simulation on a limited-size circuit and highlight the tradeoff between the level of approximation of the circuit, and the correctness of the results. We provide our benchmark scripts as a public repository~\cite{repo_thiswork}.

Our results highlight that the scalability in the runtime of an MPS simulation makes the simulation of high-qubit circuits realistic on single-GPU systems. Furthermore, we perform GPU profiling of the MPS and Tensor Network simulations. Our results show that the SVD iterations in the MPS algorithm can drastically decrease the contraction time. However, for low-qubit circuits, those SVD iterations under-utilize the GPU. Our main contributions are summarized as follows:

\begin{itemize}
    \item We develop a suite of five quantum circuits in the CUDA-Q framework;
    \item We evaluate a GPU-accelerated MPS simulator on a Grace Hopper Superchip system;
    \item We analytically model the scalability of state vector, tensor network, and MPS methods for high-qubit circuits;
    \item We propose and apply a method for assessing the correctness of MPS simulation, at various levels of circuit approximation;
    \item Our results highlight the feasibility of MPS simulation for high-qubit circuits on single-GPU systems.
\end{itemize}

\section{Background}
Quantum simulators offers scientists with the feasibility of simulating quantum circuits on classical hardware in the wait for availability of real quantum hardware under development. State vector (SV) simulators work by explicitly simulating the wave function (state vector) of a quantum system. This method provides a highly accurate representation of quantum states. SV quantum simulators store the quantum state of a $N$ qubits quantum system as a $2^N$-value vector in main memory. Currently, for small-scale quantum circuit simulations, SV simulators are the mostly widely used method for its simplicity in implementation and high fidelity. %

\begin{figure}[ht]
    \centering
    \includegraphics[width=0.9\linewidth]{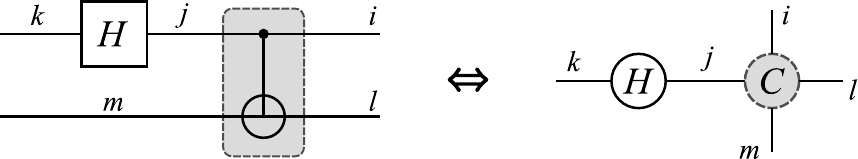}
    \caption{A quantum circuit creating an entangled bell state (left), and its equivalent representation as a tensor network (right). Adapted from~\cite{biamonte_tensor_2017}.}
    \label{fig:quantum_circuit_to_tn}
\end{figure}

A more efficient and compact way to represent a quantum state is to formulate the quantum state as factorized low-rank tensors. The factorization can be approximated, e.g., only retaining a certain number of factors. A quantum circuit can be represented as a tensor network. To convert a quantum circuit to a tensor network, each unitary quantum gate is converted into a tensor/node, and each qubit in the quantum circuit is represented as an edge in the tensor network. Figure~\ref{fig:quantum_circuit_to_tn} presents the conversion of a simple quantum circuit, which creates an entangled Bell state, to its equivalent tensor network.

Simulation of a quantum circuit represented as a tensor network is done by applying contractions over pairs of connected tensors in the network, resulting in a single tensor representing the full quantum state. For a network of $N$ tensors, $N-1$ contractions are required. As contraction is an associative operation, several sequences of contractions can be chosen. For example, contracting a network with three tensors $A$, $B$, $C$, can be done either through $(AB)C$ or $A(BC)$. Finding an optimal contraction path is a crucial step for efficient tensor network-based simulation.

\subsection{Matrix Product State}

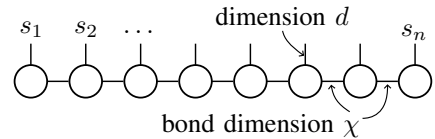
\begin{figure}[ht]
    \centering
    \def\svgwidth{0.65\linewidth}\begingroup%
  \makeatletter%
  \providecommand\color[2][]{%
    \errmessage{(Inkscape) Color is used for the text in Inkscape, but the package 'color.sty' is not loaded}%
    \renewcommand\color[2][]{}%
  }%
  \providecommand\transparent[1]{%
    \errmessage{(Inkscape) Transparency is used (non-zero) for the text in Inkscape, but the package 'transparent.sty' is not loaded}%
    \renewcommand\transparent[1]{}%
  }%
  \providecommand\rotatebox[2]{#2}%
  \newcommand*\fsize{\dimexpr\f@size pt\relax}%
  \newcommand*\lineheight[1]{\fontsize{\fsize}{#1\fsize}\selectfont}%
  \ifx\svgwidth\undefined%
    \setlength{\unitlength}{354.38436433bp}%
    \ifx\svgscale\undefined%
      \relax%
    \else%
      \setlength{\unitlength}{\unitlength * \real{\svgscale}}%
    \fi%
  \else%
    \setlength{\unitlength}{\svgwidth}%
  \fi%
  \global\let\svgwidth\undefined%
  \global\let\svgscale\undefined%
  \makeatother%
  \begin{picture}(1,0.28771022)%
    \lineheight{1}%
    \setlength\tabcolsep{0pt}%
    \put(0,0){\includegraphics[width=\unitlength,page=1]{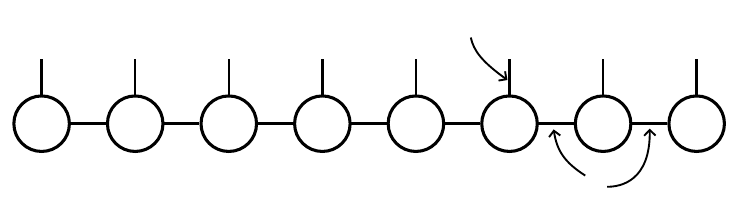}}%
    \put(0.81600538,0.00386902){\color[rgb]{0,0,0}\makebox(0,0)[rt]{\lineheight{1.25}\smash{\begin{tabular}[t]{r}bond dimension $\chi$\end{tabular}}}}%
    \put(0.05799697,0.23214951){\color[rgb]{0,0,0}\makebox(0,0)[t]{\lineheight{1.25}\smash{\begin{tabular}[t]{c}$s_1$\end{tabular}}}}%
    \put(0.18444767,0.23214951){\color[rgb]{0,0,0}\makebox(0,0)[t]{\lineheight{1.25}\smash{\begin{tabular}[t]{c}$s_2$\end{tabular}}}}%
    \put(0.3103037,0.23214951){\color[rgb]{0,0,0}\makebox(0,0)[t]{\lineheight{1.25}\smash{\begin{tabular}[t]{c}$\dots$\end{tabular}}}}%
    \put(0.94187903,0.22437029){\color[rgb]{0,0,0}\makebox(0,0)[t]{\lineheight{1.25}\smash{\begin{tabular}[t]{c}$s_n$\end{tabular}}}}%
    \put(0.63686564,0.25125294){\color[rgb]{0,0,0}\makebox(0,0)[t]{\lineheight{1.25}\smash{\begin{tabular}[t]{c}dimension $d$\end{tabular}}}}%
  \end{picture}%
\endgroup%

    \caption{Tensor network representation of a matrix product state (MPS). Each of the $n$ qubits is represented by a $s_i$.}
    \label{fig:mps}
\end{figure}

Matrix product state (MPS) is a particular case of tensor network, which can be represented as a chain of tensors. Figure~\ref{fig:mps} presents such a tensor network. To convert a general-form tensor into an MPS, a singular value decomposition (SVD) algorithm is employed recursively, starting from the original tensor representing a quantum state, to finally obtain an MPS. The obtained MPS is an exact representation of the original quantum state, and thus, the memory footprint of such representation still scales exponentially with the number of qubits. However, \ul{by truncating the number of singular values kept in each iteration of the SVD algorithm, the memory footprint can be reduced}. This number of singular values is the bond dimension $\chi$ in Figure~\ref{fig:mps}. Decreasing $\chi$ reduces the size of the representation, similarly to \emph{lossy compression} techniques. This number can be either set as a fixed number of singular values to keep or by setting a cutoff value below which singular values are eliminated. We refer the reader to~\cite{biamonte_tensor_2017} for detailed information on the MPS algorithm.

\subsection{Memory Requirement of MPS Quantum Simulators}
\label{sec:mem_footprint}
A state vector simulation requires $2^n$ complex values to represent the quantum state of a $n$-qubits system, i.e. an exponential memory usage with respect to the number of qubits. However, a recent Nvidia H100 GPU, with 96~GB GPU memory can only hold the state of a 33-qubit quantum system~\cite{schieffer2024harnessing}, even when using single precision. To simulate circuits with a higher number of qubits, multi-node multi-GPU simulations have been employed~\cite{tabuchi_mpiqulacs_2023,li_sv-sim_2021,faj2023quantum}. However, considering a multi-node, multi-GPU state vector simulation, even using the entire Frontier supercomputer with 4.7~PB of total GPU memory would only fit a 49-qubit simulation. %
The inherent exponential memory usage of state vector simulations makes it infeasible to simulate high-qubit quantum systems. Therefore, other methods must be explored for simulating large-scale quantum circuits.

For tensor network simulation, the number of parameters to represent a tensor network is $dn\chi^2$, where $n$ is the number of indices in the tensor network, i.e. the number of qubits, $d$ is the dimension, i.e. $2$ in the case of quantum circuit MPS simulation, and $\chi$ is the bond dimension. The bond dimension $\chi$ is directly linked with the degree of entanglement of the circuit. From this expression, it becomes apparent that for a set bond dimension $\chi$, the memory footprint scales linearly with the number of qubits $n$.

\section{Methodology}
In quantum computing, two components are generally relevant for the end user. First, the quantum framework provides a comprehensive set of tools, libraries, and methods to enable the development and execution of quantum circuits. We use Nvidia's CUDA-Q framework, which is still widely unexplored in the literature and provides acceleration with Nvidia GPUs.

The second component that quantum computing practitioners need to consider is the backend, which is responsible for executing or simulating quantum circuits. Each quantum framework supports various backends, either executing on quantum computer hardware that is often accessible over the cloud or simulating quantum circuits on classical computers. In this work, we focus on simulator backends that support tensor network simulations, including matrix product state (MPS) simulations on GPUs. In particular, we use the GPU-accelerated simulation backends in CUDA-Q as presented in Table~\ref{tab:cudaq_backends}.

\begin{table}[ht]
    \centering
    \caption{Backends for state vector and tensor network simulations in CUDA-Q. Names refer to backend names used when calling \texttt{cudaq.set\_target}.}
    \begin{adjustbox}{width=1\linewidth}
    \begin{tabular}{l|l|c|c}
         \hline
         Simulator name & Method & GPU & Underlying library \\
         \hline\hline
         tensornet      & Exact Tensor Network  & \cmark & cuTensorNet~\cite{cuquantum} \\
         tensornet-mps  & Matrix Product State & \cmark & cuTensorNet~\cite{cuquantum} \\
         nvidia         & State vector & \cmark & cuStateVec~\cite{cuquantum} \\
         qpp-cpu       & State vector & \xmark & Q\texttt{++}~\cite{qpp_lib} \\\hline
    \end{tabular}
    \end{adjustbox}
    \label{tab:cudaq_backends}
\end{table}

\subsection{Quantum Framework}
We use the CUDA-Q Quantum Framework in this work. Nvidia CUDA Quantum (CUDA-Q) is a set of libraries and tools providing support for developing and executing quantum circuit simulations, specifically targeting GPUs. CUDA-Q also provides support for hybrid quantum-classical algorithms. It includes a Python library for developing quantum circuits and algorithms, a compiler (\texttt{nvq++}), and a runtime to execute quantum circuits. The simulators used in this work rely on two libraries, cuStateVec and cuTensorNet. cuStateVec accelerates state vector simulations, while cuTensorNet provides GPU-accelerated algorithms for tensor network simulations. The two libraries expose a set of APIs that developers can use in their own simulators to accelerate simulations on GPUs.

In this work, we use the simulation framework that support both state vector simulations and tensor network simulations. Table~\ref{tab:cudaq_backends} summarizes the four simulators from CUDA-Q that are considered and evaluated in this work. Note that the \textit{qpp-cpu} simulator does not support GPU acceleration, and thus, is presented for completeness. As described in the table, each simulator relies on a specialized underlying library, i.e., either cuStateVec or cuTensorNet, depending on the method. This is similar to other popular simulators like Qiskit Aer~\cite{qiskit_aer}. Furthermore, part of the cuTensorNet library relies on the ExaTN library~\cite{lyakh2022exatn}.

In the MPS method, additional parameters are defined to control which singular values are kept when iterating the SVD algorithm. These parameters control the level of approximation of the original tensor network. These parameters are set via the environment variables \verb|CUDAQ_MPS_MAX_BOND|, which sets the maximum number of singular values to keep, \verb|CUDAQ_MPS_ABS_CUTOFF|, which sets a cutoff value, and \verb|CUDAQ_MPS_RELATIVE_CUTOFF|, which sets a cutoff value as a ratio. In our experiments, we use the default values of $64$, $10^{-5}$, and $10^{-5}$, respectively, for those three parameters. The SVD algorithm in use is set with the environment variable \verb|CUDAQ_MPS_SVD_ALGO|. We use the default QR algorithm for this work.

\subsection{Quantum Circuits}
We identify five quantum circuits (Table~\ref{tab:circuits}) representative of quantum algorithms that are promising for solving complex problems more efficiently than classical computing. However, they need to simulate a high number of qubits to mimic real-world problems. With state vector simulation, the memory footprint of the simulation scales exponentially with the number of simulated qubits, making unrealistic large-qubit simulations on a single-GPU system. In tensor network simulation, the memory footprint scales differently, depending on the parameters chosen for the simulation, it scales linearly with the number of qubits, rendering realistic the execution for several hundred qubits. Table~\ref{tab:circuits} summarizes the characteristics of the evaluated circuits, along with the number of gates, and the entanglement ratio.

In the context of tensor networks, the level of entanglement in a circuit is highly relevant, where lower entanglement reduces the computation cost in tensor networks and allows more approximation in the MPS method, i.e., lower memory usage. In this work, we quantify the level of entanglement using the entanglement ratio, which is defined as the ratio between the number of two-qubit gates $N_{2q}$ over the total number of gates $N_{tot}$, as introduced in~\cite{tomesh_supermarq_2022}. This value loosely characterizes the level of entanglement in the circuit, where a higher entanglement ratio indicates a higher entanglement.

\begin{table}[bt]
\centering
\caption{Quantum circuits used in our evaluation, and associated sources. $n$ is the number of qubits.}
\begin{adjustbox}{width=1\linewidth}
\begin{tabular}{l|c|c|c}
\\\hline
Name & Gates & \makecell{Entanglement \\ ratio} & Source \\
\hline\hline
Counterfeit coin & $4n-1$ & $\frac{n}{4n-1}$ & QASMBench~\cite{li_qasmbench_2023} \\
\hline
GHZ state & $n$ & $\frac{n-1}{n}$ & CUDA-Q samples~\cite{cudaq} \\
\hline
Quantum Fourier Transform & $\frac{1}{2}n(n+5)$ & $\frac{n+1}{n+5}$ & CUDA-Q samples~\cite{cudaq} \\
\hline
Quantum Volume & $\frac{1}{2}n^2$ & $1$& Qiskit Aer~\cite{qiskit_aer} \\
\hline
QAOA & $\frac{1}{2}n(2n+1)$ & $\frac{3(n-1)}{3n+1}$& SupermarQ\cite{tomesh_supermarq_2022}\\\hline
\end{tabular}
\label{tab:circuits}
\end{adjustbox}
\end{table}

\begin{figure}[bt]
    \centering
    \begin{subfigure}{\linewidth}
        \centering
        \includegraphics[width=\linewidth]{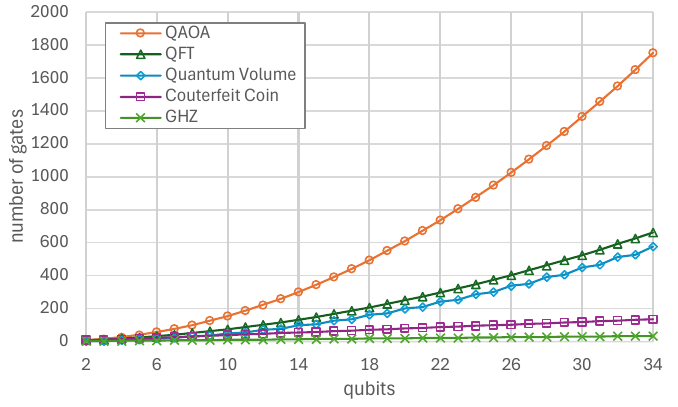}
        \caption{number of gates}
        \label{fig:gate_number}
    \end{subfigure}\hfill
    \begin{subfigure}{\linewidth}
        \centering
        \includegraphics[width=\linewidth]{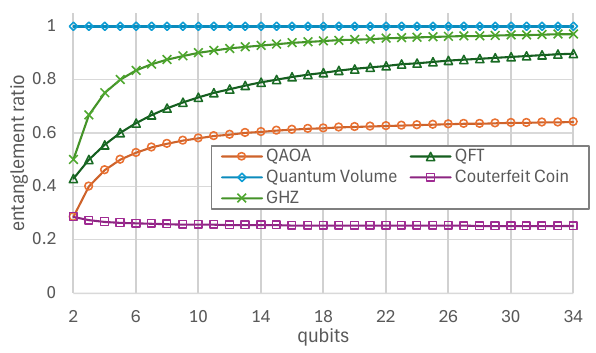}
        \caption{entanglement ratio}
        \label{fig:entanglement_ratio}
    \end{subfigure}
    \caption{The number of gates (\subref{fig:gate_number}) and entanglement ratios (\subref{fig:entanglement_ratio}) of the five evaluated circuits, for $n$ qubits. \textit{Note}: the gate count in a QFT circuit depends on the input problem, therefore, the entanglement ratio presented for QFT circuit is a lower-bound.}
\end{figure}

We analytically quantify the entanglement ratio for the five circuits at an increased number of qubits, as presented in Figure~\ref{fig:entanglement_ratio}. Asymptotically, the circuits exhibit a constant entanglement ratio, with values varying ranging from $0.25$ for Counterfeit Coin, and up to $1.0$ for Quantum Volume. This property would allow us to correlate this metric with tensor network simulation performance.

\subsubsection{Quantum Circuits Implementation in CUDA-Q}
\label{sec:circuits_cudaq}
CUDA-Q defines the concept of \textit{quantum kernels}, which provide a way to represent quantum circuits. The programmer can write quantum kernels using either Python or C\texttt{++} style. Listing~\ref{lst:kernel_python} and Listing~\ref{lst:kernel_cpp} present those two options to define a quantum circuit. While the presented code looks similar to a regular Python/C\texttt{++} code, code inside CUDA-Q kernel has strict requirements, which makes the use of some common language constructions impossible. Such requirements are in place to allow the compilation of the CUDA-Q kernels to Quantum Intermediate Representation (QIR).

An illustrative example is that arguments passed to a kernel can only be of specific types, and the type of each argument must be annotated in the associated Python function's declaration (\texttt{def\dots}). In our example, the \texttt{qubit\_count} argument is of type \texttt{int}. Other supported types are \texttt{int}, \texttt{bool}, \texttt{float}, \texttt{complex}, along with lists of those types. Furthermore, dynamic allocation is not allowed inside kernels. Such requirements are not enforced in regular Python code. 

C\texttt{++}-defined kernels must also follow similar requirements, however, since C\texttt{++} is a compiled language and is by nature less permissive than Python, those requirements are more natural to follow for the programmer.

\noindent\begin{minipage}{0.46\linewidth}
\begin{lstlisting}[language=Python,label={lst:kernel_python},caption=Example of a CUDA-Q kernel in Python.]
@cudaq.kernel
def ghz(n: int):
  q = cudaq.qvector(n)

  h(q[0])
  for i in range(n-1):
    x.ctrl(q[i], q[i+1])

  mz(q)
\end{lstlisting}
\end{minipage}\hfill
\begin{minipage}{0.5\linewidth}
\begin{lstlisting}[language=C++,label={lst:kernel_cpp},caption=Example of a CUDA-Q kernel in C\texttt{++}.]
__qpu__ void ghz(int n) {
  cudaq::qvector q(n);

  h(q[0]);
  for(int i=0; i<n-1; i++)
    x.ctrl(q[i], q[i+1]);

  mz(q); 
}
\end{lstlisting}
\end{minipage}

Another approach to define quantum kernels is to use the just-in-time kernel creation method, where quantum kernels are built at runtime. This approach is similar to the one used in Qiskit or Cirq, where a Quantum circuit is defined as an object, and methods of this object are used to apply gates to qubits, through a Python or C\texttt{++} API. This method relaxes some constraints enforced when using compile-time kernel creation, such as the limited set of allowed types in kernel-definition code. 

As CUDA-Q does not provide a direct way of importing circuits as, e.g., QASM files, we manually convert the five circuits into CUDA-Q kernels. This is done by transliterating QASM- or Qiskit-expressed circuits to CUDA-Q kernels. The code is available as a public repository~\cite{repo_thiswork}.

\subsection{Programmability}
In this work, we first aimed at leveraging Qiskit as a frontend, to import and develop quantum circuits, and use CUDA-Q simulators as a backend. As detailed in Section~\ref{sec:circuits_cudaq}, the CUDA-Q framework uses a specific programming interface to express quantum circuits -- CUDA-Q kernels. A proposed approach to interface CUDA-Q with other quantum frameworks such as Qiskit, is to express quantum circuits in the Quantum Intermediate Representation (QIR). CUDA-Q simulators can execute QIR-expressed circuits. However, as this option is not provided out-of-the-box, we exclude it from our study.

\subsection{Evaluation Platform}
Our testbed is the Grace Hopper Superchip, which features an Arm CPU with 72 Neoverse N2 cores, along with a Hopper H100 GPU. The memory on this system is 480~GB LPDDR5X for the CPU and 96~GB HBM3 for the GPU. In our experiments, quantum circuit simulations are performed as 1024-shot sampling, using the \verb|cudaq.sample| function, for the three backends, i.e. state vector, exact tensor network, and tensor network (MPS). A simulation is executed one time for warm-up and is then repeated ten times, recreating the quantum circuit at each iteration. %

\section{Results}
\label{sec:results}

\begin{figure*}[ht]
    \centering

    \begin{subfigure}{.25\linewidth}
        \centering
        \includegraphics[width=\linewidth]{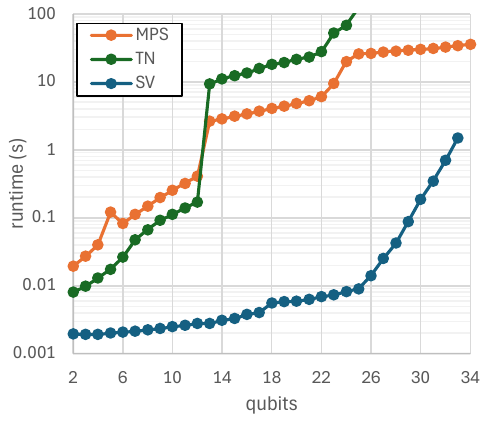}
        \caption{QFT}
        \label{fig:runtime_qft}
    \end{subfigure}\hfill
    \begin{subfigure}{.25\linewidth}
        \centering
        \includegraphics[width=\linewidth]{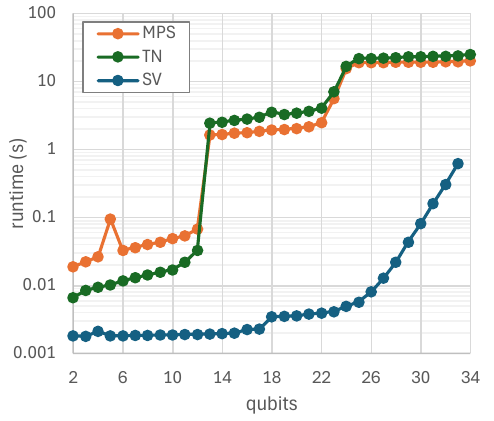}
        \caption{GHZ}
        \label{fig:runtime_ghz}
    \end{subfigure}\hfill
    \begin{subfigure}{.25\linewidth}
        \centering
        \includegraphics[width=\linewidth]{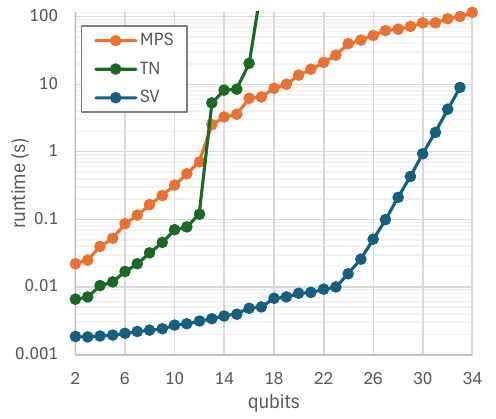}
        \caption{QV}
        \label{fig:runtime_qv}
    \end{subfigure}\hfill
    \begin{subfigure}{.25\linewidth}
        \centering
        \includegraphics[width=\linewidth]{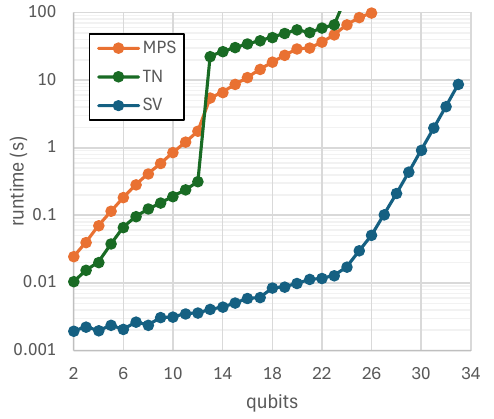}
        \caption{QAOA}
        \label{fig:runtime_qaoa}
    \end{subfigure}
    
    \caption{Runtime for increasing number of qubits for each circuit, simulation using state vector (SV), approximate matrix product state (MPS), and exact tensor network (TN).}
    \label{fig:runtime_all}
\end{figure*}

\subsection{Runtime for Low-qubits Circuits}
We measure the runtime when increasing the number of qubits for our five circuits. We use the CUDA-Q state vector simulator as a baseline for comparison. On our setup, with \qty{96}{\giga\byte} GPU memory, we can accommodate a 33-qubit single-precision state vector, running this simulation for a higher number of qubits with CUDA-Q fails.

Figure~\ref{fig:runtime_all} presents the runtime of 1024-shot simulation of four quantum circuits: Quantum Fourier Transform (QFT), GHZ state (GHZ), Quantum Volume (QV), and QAOA. For all four circuits, in configurations where state vector simulation can be used, the state vector simulation consistently outperforms tensor network and MPS simulation. This is expected for two reasons. First, in tensor network simulation, each shot corresponds to a full execution of a series of tensor contractions, while for state vector simulation, once the state vector has been computed, which is an exponential cost of the number of qubits, multiple shots can be simulated with limited per-shot cost, compared to tensor network contractions. In addition, tensor networks are used in the first place for their ability to trade a lower memory usage for a higher computational cost compared to state vector. Therefore, using tensor networks as a replacement when state vector simulation is viable is not beneficial, which we demonstrate here.

However, we observe that the scaling of the runtime is different between tensor network simulations and state vector. State vector runtime scales exponentially with the number of qubits, while tensor network simulation presents a weaker scaling. We further analyze this scaling behavior in Section~\ref{sec:modelling}.

Comparing the two tensor network methods, we observe that exact tensor network simulation exhibits a lower execution time than the MPS alternative for number of qubits below 12, after which the MPS method exhibits a lower runtime. We suggest that this is due to the overhead of executing the singular values decomposition in MPS, which improves contraction performance for the resulting tensor network, but is not beneficial when the exact tensor network is small, compared to performing contractions on the full network.

From the runtime measurements, for both tensor network and MPS simulation, we observe a significant gap in performance from 12 to 13 qubits. We suggest that this is either (1) the size of the problem, e.g. it can fit in cache below 12 qubits; or (2) a change of behavior of the simulator, where various techniques are used based on circuit characteristics; we were not able to identify such behavior in the documentation.

For the counterfeit coin (CC) circuit, which is adapted from QASMBench~\cite{li_qasmbench_2023}, we evaluated the original 12-qubits circuit. In this configuration, the reference state vector simulation takes \qty{2.8}{\second} to execute, which is significantly higher than any of the other four circuits. Similarly, for tensor network and MPS simulation, the execution time for this 12-qubit circuit is \qty{4}{\minute} and \qty{19}{\minute}, respectively. This high execution time does not allow any simulation for significantly higher number of qubits. We suggest that this is due to the structure of the circuit, where intermediary measurements are performed, and where quantum gates are applied depending on the result of those measurements. Such conditional structure drastically slows down the simulation, for both state vector and tensor network/MPS simulation.

\subsection{High-qubits Simulation}
\label{sec:modelling}
Tensor network methods enable simulations that would be otherwise impossible to execute with state vector method. In our system configuration, this corresponds to any simulation with more than 33 qubits. Above this limit, state vector simulation exceeds the memory of a single GPU, and therefore requires additional GPUs, creating a need for inter-GPU communication. Furthermore, we note here that for single-precision state vector simulation, without dedicated optimization, a 50-qubit state vector would require 8~PB of memory.
In the case of MPS, circuits with a high number of qubits can be simulated on a single GPU. In our experiments, we are able to reach 60 qubits with MPS, for all four evaluated circuits, and 90 qubits in the case of the GHZ circuit.

\begin{figure}
    \centering
    \begin{subfigure}{\linewidth}

        \includegraphics[width=\linewidth]{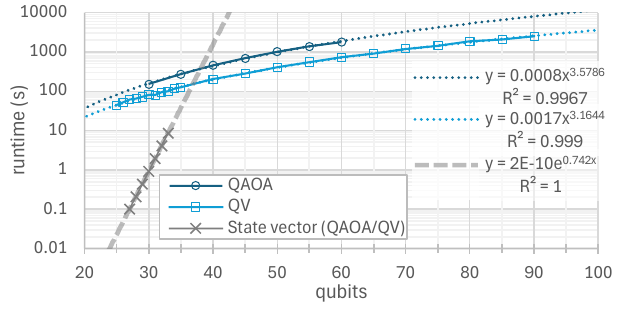}
        \caption{Power scaling}
        \label{fig:model-a}
    \end{subfigure}\hfill
    \begin{subfigure}{\linewidth}
        \centering

        \includegraphics[width=\linewidth]{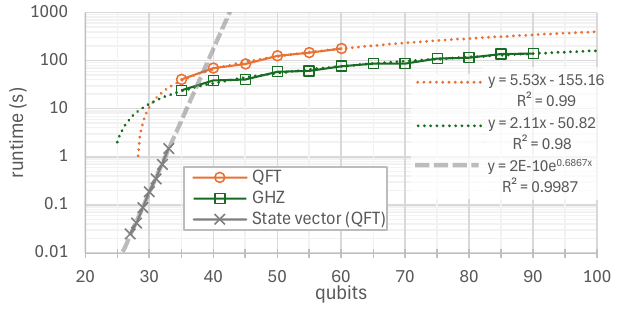}
        \caption{Linear scaling}
        \label{fig:model-b}
    \end{subfigure}\hfill
    \caption{Runtime measurements for the four circuits, with qubit count $n\geq35$, along with model (dashed line). Results are split based on the scaling behavior of the runtime with respect to the number of qubits $n$; either power ($t=\alpha n^\beta$, top) or linear ($t=an+b$, bottom). The projected runtime for state vector simulation is plotted, in the unrealistic hypothesis of infinite memory on a single GPU.}
    \label{fig:model}
\end{figure}

Figure~\ref{fig:model} presents the runtime for the four circuits, with a number of qubits $n\geq 35$. We split the results based on the scaling behavior of the runtime with respect to the number of qubits. On Figure~\ref{fig:model-a}, runtime for QAOA and QV both exhibit a runtime scaling as a power of the number of qubits $n$, in the form $t=\alpha.n^\beta$, with $\alpha$, $\beta$ constants. Figure~\ref{fig:model-b} shows that the runtime for both QFT and GHZ scales linearly with respect to the number of qubits, that is, $t=an+b$ with $a$, $b$ constants. In both cases -- linear and power scaling~--, the scaling is weaker than the exponential scaling of the runtime for state vector simulation, enabling simulation of high-qubits circuits.

On Figure~\ref{fig:model}, we also present the runtime scaling behavior of state vector simulation, in the hypothesis of infinite GPU memory. While this hypothesis is highly unrealistic, it allows to compare the exponential computational cost of state vector simulation with MPS simulation. Furthermore, we show that MPS simulation is able to simulate 60-qubit circuits in all evaluated scenarios on a single GPU, whereas state vector simulation would not be able to execute any of those simulations on any currently existing system.

Among the evaluated circuits, we observed two fundamentally different relationships between the number of qubits and the runtime of MPS simulation. This can be expected, as in contrast with state vector simulation, the computational cost of MPS simulation does not directly depend on the number of qubits. Instead, it is influenced by the structure of the circuit. This structure can be summarized as, for example, the number of gates, level of entanglement, depth, or width. %

\subsection{Profiling}

To obtain a preliminary understanding of the efficiency and bottlenecks of tensor network simulations on GPU, we perform profiling on the application. For this purpose, we chose a limited problem size, namely a Quantum Fourier Transform (QFT) circuit, with 20 qubits and 10 shots, and we compared tensor network and MPS approaches for one warmup iteration and one iteration. This small-scale setup allows to acquire reasonably-sized profiling reports. We execute the unmodified Python code in Nvidia Nsight System, using the \verb|--gpu-metric-device=0| flag to obtain statistics on GPU utilization. We further use NVTX annotations in the Python code to identify the core computational stages in the profiling report, namely the sampling of the quantum circuit kernel.

The intent of this profiling is (1) to understand the distribution of computation time between CPU and GPU and in the various computation phases, notably SVD and tensor contractions, and (2) identify, in the GPU-offloaded computations, the distribution of time between kernel execution and CPU-GPU memory movements.

\subsubsection{Overview}

\begin{figure}[t]
    \centering
    \begin{subfigure}{\linewidth}
        \centering
        \includegraphics[width=\linewidth]{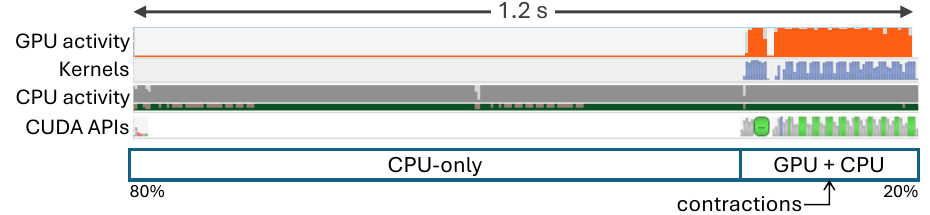}
        \caption{Exact Tensor Network}
        \label{fig:overall_profiling_tn}
        \vspace{0.5em}
    \end{subfigure}\hfill
    \begin{subfigure}{\linewidth}
        \centering
        \includegraphics[width=\linewidth]{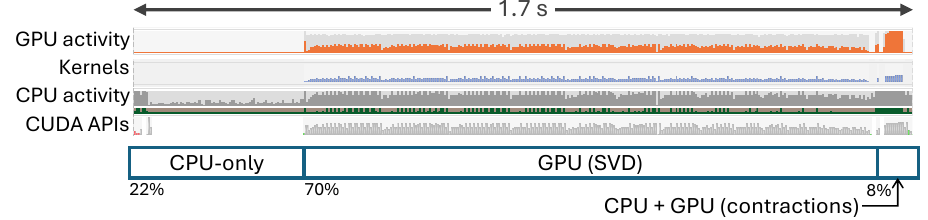}
        \caption{MPS}
        \label{fig:overall_profiling_tnmps}
    \end{subfigure}
    \caption{Profiling of the computational phase, obtained with Nvidia Nsight System, for a 10-shot simulation of a 20 qubits QFT circuit, for Tensor Network (\subref{fig:overall_profiling_tn}) and MPS (\subref{fig:overall_profiling_tnmps}). ``GPU activity'' reports the percentage of GPU cycles when the compute pipeline is non-empty.}
    \label{fig:overall_profiling}
\end{figure}

Figure~\ref{fig:overall_profiling_tn} presents the results of the profiling obtained with Nsight Systems on the overall execution of one 10-shot sampling, for exact tensor network simulation. We can divide the overall simulation into two phases. First, a CPU-only phase, representing 80\% of the total simulation time. This first phase might be linked with the preparation of the tensor network, which is performed on the CPU. This phase is followed by a second phase, where both CPU and GPU are active, representing the 20\% remaining time. From the kernel names reported in Nsight Systems, this phase appears to be the contraction part. 

Similarly, Figure~\ref{fig:overall_profiling_tnmps} reports the profiling results for the MPS simulation, the two phases previously observed remain, representing 22\% and 8\% of the execution time. However, the dominant phase of the computation in MPS simulation lies in a GPU-only phase, corresponding to iterations of the SVD algorithm, and represents 70\% of the simulation time. We note that this SVD phase only partially utilizes the GPU, as the average activity reported in the profiling is 33\%. This might indicate that the problem is too small to leverage available GPU resources. %

\subsubsection{SVD and Tensor Contractions}

\begin{figure}[ht]
    \centering
    \begin{subfigure}{\linewidth}
        \centering
        \includegraphics[width=\linewidth]{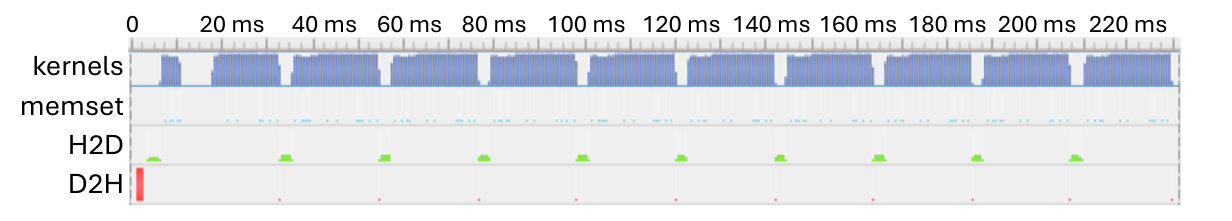}
        \caption{Exact Tensor Network}
        \label{fig:tn_contractions_profile}
    \end{subfigure}\hfill
    \begin{subfigure}{\linewidth}
        \centering
        \includegraphics[width=\linewidth]{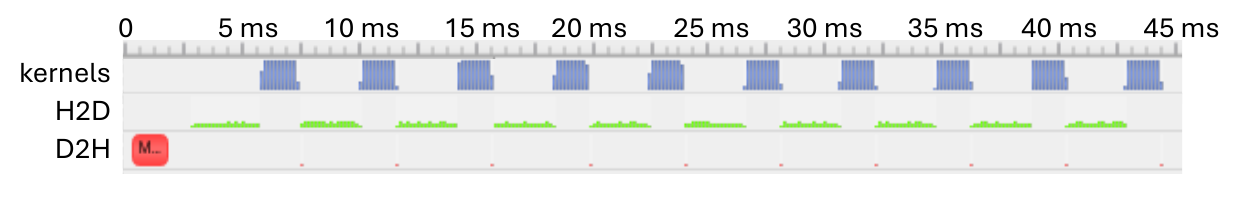}
        \caption{MPS}
        \label{fig:tnmps_contractions_profile}
    \end{subfigure}
    \caption{Profiling of the contraction part of the quantum circuit simulation, obtained with Nvidia Nsight System for a 10-shot simulation of a 20 qubits QFT circuit, for Exact Tensor Network (\subref{fig:tn_contractions_profile}) and MPS (\subref{fig:tnmps_contractions_profile}). ``H2D'' and ``D2H'' rows represent host-to-device and device-to-host data movements, respectively.}
    \label{fig:contractions_profile}
\end{figure}

Figure~\ref{fig:tn_contractions_profile} and Figure~\ref{fig:tnmps_contractions_profile} present the results of the profiling for the contraction step in the simulation, for Tensor Network and MPS simulations, respectively. This step takes \qty{225}{\milli\second} for tensor network, and \qty{45}{\milli\second} for MPS. It is important to note here that those durations would increase linearly with the number of shots, making optimization of this step critical, even for MPS where the runtime is relatively small compared to the SVD iterations. Those measurements highlight the effectiveness of SVD decomposition in MPS to reduce the computational cost of performing the contractions, induced by the simplification of the tensor network.

In the profiling results for MPS, we observe that no kernels are running for a significant part of the time (60\%), during which host-to-device memory movements are performed, with small data transfers, on the order of 128~bytes per operation. This suggests that CPU processing is performed in connection with those data movements, which could task inherently non-parallelization. While our profiling cannot indicate exactly the role of those CPU-side computations, we suggest that the task executed is ill-suited for GPU parallelization.

Interestingly, Nsight Systems reports a utilization of Tensor Cores below 1\% for both methods, over the whole execution. This is surprising, as matrix operations performed both for SVD iterations and tensor contractions can typically leverage Tensor Cores.

\section{Approximation and Correctness}
In Section~\ref{sec:results}, we demonstrated how the MPS-based method can simulate quantum circuits with a high number of qubits, where state vector simulation could not be employed. However, this comes at the cost of relying on an approximation of the quantum circuit, which might lead to incorrect results. In particular, the bond dimension $\chi$, which controls the level of approximation in the MPS simulation, is expected to influence the correctness of the output. Therefore, the environment variables \texttt{CUDAQ\_MPS\_MAX\_BOND}, \texttt{CUDAQ\_MPS\_ABS\_CUTOFF} and \texttt{CUDAQ\_MPS\_RELATIVE\_CUTOFF}, which constraint the value of $\chi$, are of interest for the user.

In this section, we focus on a single 10-qubit QAOA circuit. First, we compare the results of MPS simulation with those of state vector simulation, with the default MPS parameters. In a second part, we evaluate the impact of increasing the level of approximation in the MPS method and identify how simulation results are affected.

\subsection{Result Validation}
\label{sec:validation}
\begin{figure}[ht]
    \centering
    \includegraphics[width=\linewidth]{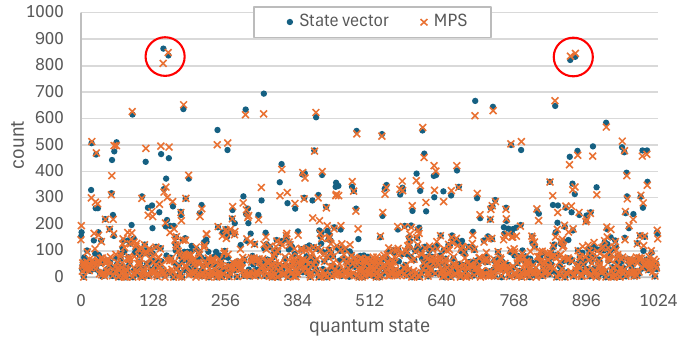}
    \caption{Distribution of measurements for a 100,000-shot simulation of a 10-qubit QAOA circuit, for state vector and MPS. Quantum states are presented in their decimal representation. For example, the state \texttt{0100100010} is represented as 290 on the x-axis. The four most likely outcomes are circled in red.}
    \label{fig:qaoa_histo}
\end{figure}

When sampling the state vector for a 10-qubit circuit, the number of possible outcomes is $2^{10}=1024$. To collect significant results, which can be compared between state vector and MPS simulation, we perform sampling with 100,000 shots. Figure~\ref{fig:qaoa_histo} presents the observation count for each state in a 100,000-shot simulation of the QAOA circuit, using either state vector or MPS. On this graph, we observe that the four most likely outcomes, circled in red, are preserved across both methods. This is a key property, as quantum algorithms often provide an output as an observed state produced with a high probability when sampling a given quantum circuit. This is the case for QAOA.

In the overall QAOA algorithm, the parameters of a quantum circuit, similar to the circuit that we present isolated here, are tuned using classical optimization. At the end of the training, the result of the problem is obtained by sampling the circuit using a set of fixed input parameters, tied to the studied problem. The most likely output is taken as the solution for the problem. In our case, the QAOA circuit is initialized with random parameters, and the measurements obtained with sampling itself do not hold any problem-specific meaning. However, we demonstrate that in this simple setup, MPS-based simulation preserves the most likely outcomes, when compared to reference exact state vector simulation.

\subsection{Impact of MPS Approximation}

\begin{table}[ht]
    \newcommand{\G}{\cellcolor{green!25}}
    \newcommand{\R}{\cellcolor{red!25}}
    \renewcommand{\S}[1]{\texttt{#1}}

    \newcolumntype{Y}{>{\centering\arraybackslash}X}
    
    \centering
    \caption{Four most sampled quantum states in a 100,000-shot sampling of a QAOA circuit, with 10 qubits. Each state is represented in its decimal form. Green color indicates a match with the list of states obtained from a reference state vector simulation (column ``SV'').}
    \setlength\tabcolsep{2pt} 
    \begin{tabularx}{\linewidth}{|c|Y|YYYYYYYY|}
         \hline
            \multirowcell{2}{rank} & \multirowcell{2}{SV \\ (\textit{ref.})} & \multicolumn{8}{c|}{MPS, $\chi_\text{max}$ \rotatebox[origin=b]{-90}{$\Rsh$} } \\

          & & \footnotesize $64$ & $32$ & $16$
              & \footnotesize $15$ & $14$ & $13$ & $12$ & $84$ \\
         \hline
         \#1 & \G \S{146} & \G \S{155} & \G \S{155} & \G \S{868}
             & \G \S{155} & \G \S{155} & \R \S{731} & \R \S{699} & \R \S{94}\\
         \#2 & \G \S{155} & \G \S{877} & \G \S{868} & \G \S{155}
             & \G \S{868} & \G \S{868} & \R \S{292} & \R \S{324} & \R \S{929} \\
         \#3 & \G \S{877} & \G \S{868} & \G \S{877} & \G \S{877}
             & \R \S{417} & \R \S{731} & \G \S{868} & \R \S{731} & \G \S{868} \\
         \#4 & \G \S{868} & \G \S{146} & \G \S{146} & \G \S{146}
             & \R \S{606} & \R \S{609} & \G \S{155} & \G \S{868} & \G \S{155} \\
         \hline
    \end{tabularx}
    \label{tab:four_states}
\end{table}

The environment variable \texttt{CUDAQ\_MPS\_MAX\_BOND} sets a higher bound $\chi_\text{max}$ for the bond dimension $\chi$, that is, $\chi\leq\chi_\text{max}$. This variable controls the level of approximation in the MPS simulation, with higher values providing a more accurate MPS representation of the circuit. Lower values of $\chi_\text{max}$, on the other hand, provide a more compact MPS representation of the quantum circuit and therefore reduce the computational cost and memory footprint. However, as this is an approximation, the distribution of sampled states produced by the MPS simulation with low values of $\chi_\text{max}$ is expected to differ from the reference distribution, obtained with high values of $\chi_\text{max}$, or state vector simulation.

To assess the correctness of the simulation at various levels of approximation, that is, for various values of the maximum bond dimension $\chi_\text{max}$, we use state vector simulation as reference. Using state vector simulation in our particular instance of QAOA circuit, we note that four states exhibit a significantly higher probability compared to others. Therefore, we use those four states as a reference output to assess the correctness of MPS simulation, when changing $\chi_\text{max}$. Note that this number of \textit{four} might be different for other circuits; this aspect is application-dependent. For our simulation of a 10-qubit QAOA circuit, with 100,000-shot sampling, Table~\ref{tab:four_states} presents the four most likely states for several values of $\chi_\text{max}$. The green color indicates states that match the reference result. We observe that the default value of $\chi_\text{max}=64$ provides the same four most likely states as observed in state vector simulation, as discussed in Section~\ref{sec:validation}. This observation holds when reducing $\chi_\text{max}$ down to $16$. However, below this value, only at most two of the four most likely states are preserved, down to $\chi_\text{max}=8$. In such cases, the result can be considered incorrect, since it does not reproduce the expected output obtained with state vector simulation.

It is important to note that in this experiment, we evaluate the QAOA circuit by itself, without including it in a wider use case. In a real-world scenario, this circuit is used as part of a QAOA optimization algorithm, where the parameters of the circuit are tuned iteratively. Therefore, low values of $\chi_\text{max}$, despite producing results that diverge from the reference state vector simulation, MPS simulation can still be relevant in the context of the entire QAOA algorithm, where the approximation in circuit simulation is counterbalanced by a larger number of iterations.

\section{Related works}
Various optimizations to state vector simulation have been proposed. In the context of GPU-accelerated simulation, several techniques have been proposed to tackle the issue of the memory footprint. These include cache blocking technique~\cite{doi_cache_2020}, data compression~\cite{zhang_memqsim_2023}, large shared memory space~\cite{li_sv-sim_2021,doi_quantum_2019}. Other works have been conducted to improve the performance of state vector simulation code. 
This includes optimization for long-vector architectures~\cite{takahashi_prototype_2023}, leveraging the sparsity of the state vector~\cite{chundury_diaq_2024}. However, the exponential memory footprint, inherent to state vector simulation, hinders the ability to simulate high-qubits circuits~\citet{faj_quantum_2023}. Our work aims at identifying to what extent tensor network-based methods can be used to bridge the gap in high-qubits simulation.

Tensor Network methods have been shown to provide significant opportunities to execute relevant quantum circuits, with high qubit counts. \citet{chen_cutn-qsvm_2024} execute quantum machine learning workload using tensor network simulation, with 784~qubits. \citet{nguyen_tensor_2022} were able to perform noisy simulation on circuits used to evaluate Google's 53-qubit Sycamore chip. 

Identification of circuits that would benefit from tensor network simulation is an active topic of research. Some algorithms, such as hybrid quantum-classical algorithms~\cite{mccaskey_validating_2018}, including variational quantum eigensolvers (VQE)~\cite{huang_tensor-network-assisted_2023}, have been shown to be relevant candidates. A low level of entanglement is a frequently listed characteristic~\cite{vidal_efficient_2003,orus_tensor_2019}. \citet{vallero_state_2024} evaluate the relationship between the performance of tensor network-based simulation and some key characteristics of quantum circuits. Our work aims to provide some understanding of the performance of tensor network-based methods for various quantum circuits.

\section{Conclusion and Future Work}

In this work, we explored and evaluated tensor network-based simulations of large-scale quantum circuits. We evaluated the performance of two tensor network-based methods -- exact tensor network and approximate MPS -- implemented as GPU-accelerated simulators in Nvidia's CUDA-Q framework. Our results show that MPS simulations can simulate circuits of a higher number of qubits than what is feasible in state vector simulation on the same machine. By profiling the MPS and tensor network simulations, we showed that the SVD iterations in the MPS algorithm can drastically decrease the contraction time. However, for a low-qubit circuit, those SVD iterations appear to under-utilize GPU. We further proposed an approach to identify how the bond dimension limit $\chi_\text{max}$ affects the correctness of MPS simulation and demonstrated this method on a QAOA circuit. Our study focuses on simulating quantum circuits standalone. In future work, we will include the evaluation of MPS-simulated circuits within real-world scenarios, such as part of a full QAOA algorithm. We suggest that expressing quantum circuits in the Quantum Intermediate Representation (QIR) could facilitate this integration, by enabling the use of CUDA-Q backends within pipelines comprising other tools, such as Qiskit.

\section*{Acknowledgements}
This work is funded by the European Union. This work has received funding from the European High Performance Computing Joint Undertaking (JU) and Sweden, Finland, Germany, Greece, France, Slovenia, Spain, and the Czech Republic under grant agreement No 101093261, Plasma-PEPSC.

\bibliographystyle{IEEEtranN}
\bibliography{main,quantum}

\begin{thebibliography}{32}
\providecommand{\natexlab}[1]{#1}
\providecommand{\url}[1]{#1}
\csname url@samestyle\endcsname
\providecommand{\newblock}{\relax}
\providecommand{\bibinfo}[2]{#2}
\providecommand{\BIBentrySTDinterwordspacing}{\spaceskip=0pt\relax}
\providecommand{\BIBentryALTinterwordstretchfactor}{4}
\providecommand{\BIBentryALTinterwordspacing}{\spaceskip=\fontdimen2\font plus
\BIBentryALTinterwordstretchfactor\fontdimen3\font minus \fontdimen4\font\relax}
\providecommand{\BIBforeignlanguage}[2]{{%
\expandafter\ifx\csname l@#1\endcsname\relax
\typeout{** WARNING: IEEEtranN.bst: No hyphenation pattern has been}%
\typeout{** loaded for the language `#1'. Using the pattern for}%
\typeout{** the default language instead.}%
\else
\language=\csname l@#1\endcsname
\fi
#2}}
\providecommand{\BIBdecl}{\relax}
\BIBdecl

\bibitem[Pan and Zhang(2021)]{pan_simulating_2021}
F.~Pan and P.~Zhang, ``Simulating the {Sycamore} quantum supremacy circuits,'' Mar. 2021, arXiv:2103.03074.

\bibitem[Tabuchi et~al.(2023)Tabuchi, Imamura, Yamazaki, Honda, Kasagi, Nakao, Fukumoto, and Nakashima]{tabuchi_mpiqulacs_2023}
A.~Tabuchi, S.~Imamura, M.~Yamazaki, T.~Honda, A.~Kasagi, H.~Nakao, N.~Fukumoto, and K.~Nakashima, ``{mpiQulacs}: {A} {Scalable} {Distributed} {Quantum} {Computer} {Simulator} for {ARM}-based {Clusters},'' in \emph{2023 {IEEE} {International} {Conference} on {Quantum} {Computing} and {Engineering} ({QCE})}, vol.~01, Sep. 2023, pp. 959--969.

\bibitem[Markidis(2024)]{markidis2024quantum}
S.~Markidis, ``What is quantum parallelism, anyhow?'' in \emph{ISC High Performance 2024 Research Paper Proceedings (39th International Conference)}.\hskip 1em plus 0.5em minus 0.4em\relax Prometeus GmbH, 2024, pp. 1--12.

\bibitem[Chen et~al.(2024)Chen, Li, Wang, See, Wang, Wille, Chen, Yang, and Lin]{chen_cutn-qsvm_2024}
K.-C. Chen, T.-Y. Li, Y.-Y. Wang, S.~See, C.-C. Wang, R.~Wille, N.-Y. Chen, A.-C. Yang, and C.-Y. Lin, ``{cuTN}-{QSVM}: {cuTensorNet}-accelerated {Quantum} {Support} {Vector} {Machine} with {cuQuantum} {SDK},'' May 2024, arXiv:2405.02630 [quant-ph].

\bibitem[Nguyen et~al.(2022)Nguyen, Lyakh, Dumitrescu, Clark, Larkin, and McCaskey]{nguyen_tensor_2022}
T.~Nguyen, D.~Lyakh, E.~Dumitrescu, D.~Clark, J.~Larkin, and A.~McCaskey, ``Tensor {Network} {Quantum} {Virtual} {Machine} for {Simulating} {Quantum} {Circuits} at {Exascale},'' \emph{ACM Transactions on Quantum Computing}, vol.~4, no.~1, pp. 6:1--6:21, Oct. 2022.

\bibitem[IBM()]{qiskit_aer}
IBM, ``Qiskit aer,'' \url{https://qiskit.github.io/qiskit-aer/}.

\bibitem[Google(2024)]{cirq}
Google, ``Cirq,'' \url{https://quantumai.google/cirq}, 2024.

\bibitem[Faj et~al.(2023{\natexlab{a}})Faj, Peng, Wahlgren, and Markidis]{faj2023quantum}
J.~Faj, I.~Peng, J.~Wahlgren, and S.~Markidis, ``Quantum computer simulations at warp speed: Assessing the impact of gpu acceleration: A case study with ibm qiskit aer, nvidia thrust \& cuquantum,'' in \emph{2023 IEEE 19th International Conference on e-Science (e-Science)}.\hskip 1em plus 0.5em minus 0.4em\relax IEEE, 2023, pp. 1--10.

\bibitem[Microsoft(2024)]{azure_quantum}
Microsoft, ``Azure quantum,'' \url{https://quantum.microsoft.com/}, 2024.

\bibitem[team and collaborators(2020)]{qsim}
Q.~A. team and collaborators, ``qsim,'' \url{https://doi.org/10.5281/zenodo.4023103}, 2020.

\bibitem[Markidis(2023)]{markidis2023enabling}
S.~Markidis, ``Enabling quantum computer simulations on amd gpus: a hip backend for google's qsim,'' in \emph{Proceedings of the SC'23 Workshops of The International Conference on High Performance Computing, Network, Storage, and Analysis}, 2023, pp. 1478--1486.

\bibitem[Nvidia({\natexlab{a}})]{cudaq}
Nvidia, ``Cuda-q documentation,'' \url{https://nvidia.github.io/cuda-quantum/latest/index.html}.

\bibitem[rep()]{repo_thiswork}
\url{https://github.com/KTH-ScaLab/cudaq-benchmarks/}.

\bibitem[Biamonte and Bergholm(2017)]{biamonte_tensor_2017}
J.~Biamonte and V.~Bergholm, ``Tensor {Networks} in a {Nutshell},'' Jul. 2017, arXiv:1708.00006.

\bibitem[Schieffer et~al.(2024)Schieffer, Wahlgren, Ren, Faj, and Peng]{schieffer2024harnessing}
G.~Schieffer, J.~Wahlgren, J.~Ren, J.~Faj, and I.~Peng, ``Harnessing integrated cpu-gpu system memory for hpc: a first look into grace hopper,'' in \emph{Proceedings of the 53rd International Conference on Parallel Processing}, 2024, pp. 199--209.

\bibitem[Li and et~al.(2021)]{li_sv-sim_2021}
A.~Li and et~al., ``{SV}-sim: scalable {PGAS}-based state vector simulation of quantum circuits,'' in \emph{Proceedings of the SC21}, ser. {SC} '21.\hskip 1em plus 0.5em minus 0.4em\relax New York, NY, USA: Association for Computing Machinery, Nov. 2021, pp. 1--14.

\bibitem[Nvidia({\natexlab{b}})]{cuquantum}
Nvidia, ``cuquantum sdk documentation,'' \url{https://docs.nvidia.com/cuda/cuquantum/}.

\bibitem[Gheorghiu(2018)]{qpp_lib}
V.~Gheorghiu, ``Quantum++: A modern c++ quantum computing library,'' \emph{PLOS ONE}, vol.~13, no.~12, pp. 1--27, 12 2018.

\bibitem[Lyakh et~al.(2022)Lyakh, Nguyen, Claudino, Dumitrescu, and McCaskey]{lyakh2022exatn}
D.~I. Lyakh, T.~Nguyen, D.~Claudino, E.~Dumitrescu, and A.~J. McCaskey, ``Exatn: Scalable gpu-accelerated high-performance processing of general tensor networks at exascale,'' \emph{Frontiers in Applied Mathematics and Statistics}, vol.~8, p. 838601, 2022.

\bibitem[Tomesh et~al.(2022)Tomesh, Gokhale, Omole, Ravi, Smith, Viszlai, Wu, Hardavellas, Martonosi, and Chong]{tomesh_supermarq_2022}
T.~Tomesh, P.~Gokhale, V.~Omole, G.~S. Ravi, K.~N. Smith, J.~Viszlai, X.-C. Wu, N.~Hardavellas, M.~R. Martonosi, and F.~T. Chong, ``\BIBforeignlanguage{English}{{SupermarQ}: {A} {Scalable} {Quantum} {Benchmark} {Suite}},'' in \emph{\BIBforeignlanguage{English}{Proceedings of the 2022 {IEEE} {International} {Symposium} on {High}-{Performance} {Computer} {Architecture} ({HPCA})}}.\hskip 1em plus 0.5em minus 0.4em\relax IEEE Computer Society, Apr. 2022, pp. 587--603.

\bibitem[Li et~al.(2023)Li, Stein, Krishnamoorthy, and Ang]{li_qasmbench_2023}
A.~Li, S.~Stein, S.~Krishnamoorthy, and J.~Ang, ``{QASMBench}: {A} {Low}-{Level} {Quantum} {Benchmark} {Suite} for {NISQ} {Evaluation} and {Simulation},'' \emph{ACM Transactions on Quantum Computing}, vol.~4, no.~2, pp. 10:1--10:26, 2023.

\bibitem[Doi and Horii(2020)]{doi_cache_2020}
J.~Doi and H.~Horii, ``Cache {Blocking} {Technique} to {Large} {Scale} {Quantum} {Computing} {Simulation} on {Supercomputers},'' in \emph{2020 {IEEE} {International} {Conference} on {Quantum} {Computing} and {Engineering} ({QCE})}, Oct. 2020, pp. 212--222.

\bibitem[Zhang et~al.(2023)Zhang, Fang, Guan, Li, and Tao]{zhang_memqsim_2023}
B.~Zhang, B.~Fang, Q.~Guan, A.~Li, and D.~Tao, ``\BIBforeignlanguage{en}{{MEMQSim}: {Highly} {Memory}-{Efficient} and {Modularized} {Quantum} {State}-{Vector} {Simulation}},'' in \emph{\BIBforeignlanguage{en}{Proceedings of the {SC} '23 {Workshops} of {The} {International} {Conference} on {High} {Performance} {Computing}, {Network}, {Storage}, and {Analysis}}}.\hskip 1em plus 0.5em minus 0.4em\relax ACM, Nov. 2023, pp. 1452--1453.

\bibitem[Doi et~al.(2019)Doi, Takahashi, Raymond, Imamichi, and Horii]{doi_quantum_2019}
J.~Doi, H.~Takahashi, R.~Raymond, T.~Imamichi, and H.~Horii, ``Quantum computing simulator on a heterogenous {HPC} system,'' in \emph{Proceedings of the 16th {ACM} {International} {Conference} on {Computing} {Frontiers}}, ser. {CF} '19.\hskip 1em plus 0.5em minus 0.4em\relax New York, NY, USA: Association for Computing Machinery, Apr. 2019, pp. 85--93.

\bibitem[Takahashi et~al.(2023)Takahashi, Mori, and Takizawa]{takahashi_prototype_2023}
K.~Takahashi, T.~Mori, and H.~Takizawa, ``Prototype of a {Batched} {Quantum} {Circuit} {Simulator} for the {Vector} {Engine},'' in \emph{Proceedings of the {SC} '23 {Workshops} of {The} {International} {Conference} on {High} {Performance} {Computing}, {Network}, {Storage}, and {Analysis}}, ser. {SC}-{W} '23.\hskip 1em plus 0.5em minus 0.4em\relax New York, NY, USA: Association for Computing Machinery, Nov. 2023, pp. 1499--1505.

\bibitem[Chundury et~al.(2024)Chundury, Li, Suh, and Mueller]{chundury_diaq_2024}
S.~Chundury, J.~Li, I.-S. Suh, and F.~Mueller, ``{DiaQ}: {Efficient} {State}-{Vector} {Quantum} {Simulation},'' Apr. 2024, arXiv:2405.01250 [quant-ph].

\bibitem[Faj et~al.(2023{\natexlab{b}})Faj, Peng, Wahlgren, and Markidis]{faj_quantum_2023}
J.~Faj, I.~Peng, J.~Wahlgren, and S.~Markidis, ``Quantum {Computer} {Simulations} at {Warp} {Speed}: {Assessing} the {Impact} of {GPU} {Acceleration}: {A} {Case} {Study} with {IBM} {Qiskit} {Aer}, {Nvidia} {Thrust} \& {cuQuantum},'' in \emph{2023 {IEEE} 19th {International} {Conference} on e-{Science} (e-{Science})}, Oct. 2023, pp. 1--10.

\bibitem[McCaskey et~al.(2018)McCaskey, Dumitrescu, Chen, Lyakh, and Humble]{mccaskey_validating_2018}
A.~McCaskey, E.~Dumitrescu, M.~Chen, D.~Lyakh, and T.~Humble, ``\BIBforeignlanguage{en}{Validating quantum-classical programming models with tensor network simulations},'' \emph{\BIBforeignlanguage{en}{PLOS ONE}}, vol.~13, no.~12, Dec. 2018, publisher: Public Library of Science.

\bibitem[Huang et~al.(2023)Huang, He, Zhang, Wu, Wu, and Yuan]{huang_tensor-network-assisted_2023}
J.~Huang, W.~He, Y.~Zhang, Y.~Wu, B.~Wu, and X.~Yuan, ``Tensor-network-assisted variational quantum algorithm,'' \emph{Physical Review A}, vol. 108, no.~5, p. 052407, Nov. 2023, publisher: American Physical Society.

\bibitem[Vidal(2003)]{vidal_efficient_2003}
G.~Vidal, ``Efficient {Classical} {Simulation} of {Slightly} {Entangled} {Quantum} {Computations},'' \emph{Physical Review Letters}, vol.~91, no.~14, p. 147902, Oct. 2003, publisher: American Physical Society.

\bibitem[Orús(2019)]{orus_tensor_2019}
R.~Orús, ``\BIBforeignlanguage{en}{Tensor networks for complex quantum systems},'' \emph{\BIBforeignlanguage{en}{Nature Reviews Physics}}, vol.~1, no.~9, pp. 538--550, Sep. 2019, publisher: Nature Publishing Group.

\bibitem[Vallero et~al.(2024)Vallero, Vella, and Rech]{vallero_state_2024}
M.~Vallero, F.~Vella, and P.~Rech, ``State of practice: evaluating {GPU} performance of state vector and tensor network methods,'' Jan. 2024, arXiv:2401.06188.

\end{thebibliography}

\end{document}